\documentstyle[11pt]{article}
\begin{document}
\rightline{UCTP 102.97}
\bigskip
\begin{center}
{\Large\bf
Rotating Charged Solutions to Einstein-Maxwell-ChernSimons
Theory
in 2+1 Dimensions}
\end{center}
\hspace{0.4cm}
\begin{center}
{Sharmanthie Fernando \footnote{fernando@physunc.phy.uc.edu}and
Freydoon Mansouri\footnote{mansouri@uc.edu}}\\
{\small\it Department of Physics}\\
{\small\it University of Cincinnati}\\
{\small\it Cincinnati}\\
{\small\it Ohio 45221}\\
{\small\it U.S.A.}\\

\end{center}

\begin{center}
{\bf Abstract}
\end{center}

\hspace{0.7cm}{\small We obtain a class of rotating charged 
 stationary
circularly symmetric solutions of  Einstein-Maxwell theory
coupled
to a
topological mass term for the Maxwell field. These solutions are
regular, have 
finite mass and angular momentum, and are asymptotic to the
uncharged extreme
BTZ black hole.}

\section{Introduction}
Gravity in 2+1 dimensions has received a great deal of attention
in the recent past.  The rationale for this is that it is
technically
simpler so that it may be used as a theoretical laboratory for
studying certain aspects of gravity in 3+1 dimensions. In
particular, it might shed some light on the quantum gravity
problem. The technical simplicity of the 2+1 dimensional theory
stems from the fact that in 2+1 dimensions the components of
Riemann curvature tensor can be completely expressed in terms of 
Einstein tensor. This implies that Einstein's theory is trivial
in
the absence of  matter. The space time will be flat outside the
matter sources with no propagating degrees of freedom and no
Newtonian limit to the field equations. This means that the
gravitational influence
of the matter sources will be topological in character.
For example, a point source will produce a conical structure to
the space-time \cite{des1}.

The situation changes when a cosmological term of anti-de Sitter
(AdS) variety is added to Einstein's equations. In this case, as
pointed out by
Banadoz-Teitelboim-Zanellie \cite{banados}, the field
equations admit black hole solutions with 
finite mass and finite angular momentum \cite{ban2}.
Various properties of this BTZ black hole such as its
thermodynamic, statistical, and quantum properties have been
studied extensively \cite{carlip}.
There are also a number of works in which the corresponding
charged black holes have been investigated. The first of these
was static  charged black holes constructed 
by Banadoz {\it et al.} in \cite{banados}. It was then found that
due to the
logarithmic  nature of the electromagnetic  potential, the quasi
local mass of this solution diverges \cite{eric}.
Later, a horizonless static  solution with  magnetic charge was
constructed
by Hirshmann {\it et al.} \cite{eric}. In this solution, again
the quasi local mass diverges as demonstrated by Chan
\cite{chan}.
 
A rotating charged black hole solution was found by Kamata {\it
et al.} \cite{kam}. They obtained their solution by imposing a
self (anti-self) duality condition on the electromagnetic
field. The resulting solutions were asymptotic to an extreme BTZ
black hole solution but again had diverging mass and angular
momentum \cite{chan}.

From the brief summary of some of the works in recent literature
given above,
it is clear that solutions in $2+1$ dimensions with finite mass
and angular momentum are a small minority. In this paper, we 
present a general class of solutions of
Einstein-Maxwell theory, in which the
electromagnetic potential is ``screened'' and in the simplest
case
falls as $1/r$ at large distances. This
kind of behavior for the electromagnetic potential was achieved
by introducing a topological mass term for the electormagnetic
field \cite{tem}. 
Since we begin with the coupled field equations and use standard
techniques, the issues of generality of the solutions and their
uniqueness are transparent at every stage. Moreover,
we explicitly compute
the mass and the angular momentum of our solutions
and show that they are finite. 
The plan of the paper is
as follows. In section 2 we will derive the field  equations for
the coupled
gravitational and electromagnetic fields . In section 3 we will
look
for stationary circularly symmetric solutions, first in general
and
then in a more specialized form. These solutions will still have
undetermined parameters. The quasilocal
mass
and angular momentum of the solutions are  computed in section 4. 
The properties of the rotating charged solutions are summarized
and further discussed in section 5. Section 6 is devoted to 
comparison with other works.

\section{The Equations of motion}
In odd dimensional spaces, it is possible to introduce
topologically non trivial
gauge invariant terms which give mass to the gauge fields
\cite{tem}. In  2+1 dimensions, an abelian gauge field becomes
massive if the Lagrangian is modified by a Chern Simon term

$$L =
\frac{m_p}{2}\varepsilon^{\nu\alpha\beta}F_{\alpha\beta}A_{\nu}
\eqno(2.1)$$
where the gauge field, the corresponding field strength tensor,
and the
mass for the gauge field are given, respectively, by  $A_\mu$ ,
$F_{\mu\nu} =
\bigtriangledown_{\mu}A_{\nu} -   \bigtriangledown_{\nu}A_{\mu}$, 
and $m_p$.
In this work,  we couple the Maxwell theory modified by a
topological mass term to Einstein's theory in $2+1$ dimensions
and look for its general solutions. The corresponding action can
be written as

$$ I = \int d^3x L$$
where $$L = \sqrt{|g|}\left[\frac{1}{2\pi}( R-2\Lambda )-
\frac{1}{4}g^{\mu\nu}g^{\rho\sigma}F_{\mu\rho}F_{\nu\sigma}\right
]
+ \frac{m_{p}}{2}
\varepsilon^{\nu\alpha\beta}F_{\alpha\beta}A_\nu\eqno(2.2)$$
In this expression, the Newtonian constant has been chosen to be
$\frac{1}{8}$,
$ \Lambda $ is the cosmological constant, and  $m_{p}$ is the
topological mass. We assume  $\Lambda $ to be negative. It
follows that we can write down the gravitational field equations
in the form

$$R_{\mu\nu} - \frac{1}{2} g_{\mu\nu}R + \Lambda
g_{\mu\nu} = \pi T_{\mu\nu}\eqno(2.3)$$
where the energy-momentum stress energy tensor $T_{\mu\nu}$ given
by:

$$T_{\mu\nu} = F_{\mu\rho}F_{\nu}^{\rho} - \frac{1}{4}
g_{\mu\nu}F_{\lambda\sigma}F^{\lambda\sigma}\eqno(2.4)$$
Being a topological, the Chern Simons term does not contribute to
the energy momentum tensor.
The electromagnetic field equations takes the form:

$$\partial_{\mu}(\sqrt{|g|} F^{\mu\nu}) +
m_p\varepsilon^{\nu\alpha\beta}F{_{\alpha\beta}}= 0\eqno(2.5)$$

To look for solutions, we assume that the three dimensional space
time is stationary
and circularly  symmetric, i.e., it has two commuting Killing
vectors. Writing the coordinates as the triple $(t, \phi, r)$,
the two Killing vectors can be represented as
$\frac{\partial}{\partial \phi}$ and $\frac{\partial}{\partial
t}.$
Then the line element can be put in the form

$$ds^2=  - N^2dt^2  + L^{-2}dr^2 + K^2(d\phi + N^\phi
dt)^2\eqno(2.6)$$
The functions $N,L,K$ and $N^\phi$  depend only on radial
coordinate $r$. We assume that the
electromagnetic field is also circularly symmetric, so that its
only non zero components are $E_{r}$ and $B$.

To obtain solutions, it is simplest to use the tetrad formalism
and Cartan structure equations. We transform to the orthonormal
basis

$$\theta^0  = Ndt ;\hspace{1.0cm}  \theta^1 = K (d\phi + N^\phi
dt)
; \hspace{1.0cm}  \theta^2= L^{-1}dr \eqno(2.7)$$
We will use the indices $a,b = 0,1,2$ for the orthonormal basis
and $\mu,
\nu = 0,1,2$
for the coordinate basis :
$$x^0 = t; \hspace{1.0cm}  x^1 = \phi ; \hspace{1.0cm} x^2 =
r;\eqno(2.8)$$
The non-vanishing components of the electromagnetic field tensor
in the coordinate basis are given
by 

$$F_{tr} = E ; \hspace{1.0cm}  F_{r\phi} = B; \eqno(2.9)$$
whereas in the orthonormal basis, they are given by 
$$F_{02} = \tilde{E} ; \hspace{1.0cm}  F_{21} =
\tilde{B};\eqno(2.10)$$
Being a 2-form, an arbitrary field strength tensor can written in
the two bases as

$$F = F_{ab} \theta^a \wedge \theta^b  = F_{\mu\nu}dx^{\mu}\wedge
dx^{\nu}\eqno(2.11)$$
Specialized to circularly symmetric case,these expressions reduce
to
$$F = \tilde{E}\hspace{0.3cm}\theta^0 \wedge \theta^2 + 
\tilde{B}\hspace{0.3cm} \theta^2 \wedge \theta^1 = E
\hspace{0.3cm}
dx^{0}\wedge dx^{2} + B \hspace{0.3cm} dx^{2}\wedge
dx^{1}\eqno(2.12)$$
It then follows that the components in the two bases are  related
to each other by

$$E = \frac {(\tilde{E}N - \tilde{B}KN^{\phi})} {L}$$
$$B = \frac { \tilde{B} K} {L}\eqno(2.13)$$

We will use the Cartan's method to calculate the curvature
components relative to the orthonormal basis. In this basis,  
the
line element is simply given by

$$ds^2 = \theta^{i}\theta^{j}\eta_{ij}= 
-{\theta^0}^2+{\theta^1}^2+{\theta^2}^2\eqno(2.14)$$
where the one form components $\theta^i$ is given by (2.7) and
$\eta_{ij}$
=(
- + +). Let us write the connection one form

$$\Omega^{\alpha}_{\beta} = \gamma^{\alpha}_{\rho
\beta}\theta^{\rho}\eqno(2.15)$$
with the anti symmetry property 

$$\Omega_{\alpha\beta} + \Omega_{\beta\alpha}=0\eqno(2.16)$$
Then the curvature two form in this basis is given by

$$ \tilde{R^{\alpha}_{\beta}}= \tilde{R}^{\alpha} _{\beta \mu
\nu}
\theta^{\mu}\wedge \theta^{\nu}\eqno(2.17)$$
The Cartan
structure equations will relate the connection one forms and the
curvature two forms as follows.

$$d\theta^{\alpha} + \Omega^{\alpha} _{\beta}\wedge
\theta^{\beta}=0\eqno(2.18)$$

$$\tilde{R^{\alpha}_{\beta}} =  d\Omega^{\alpha} _{\beta} + 
\Omega^{\alpha} _{\mu} \wedge 
\Omega^{\mu}_{\beta}\eqno(2.19)$$\\
Equations (2.18) and (2.19) can be used to compute all the
unknown tensor components $\gamma^{\alpha}_{\rho \beta}$ and
$\tilde{R}^{\alpha}_{\beta \mu \nu}$ in the orthonormal basis and
then the corresponding expressions in the parameterization given
by equation (2.6).
Then, after computing the components of Ricci and Einstein
tensors, one can 
write down the gravitational field equation in the  following 
form :
 
$$\tilde{G_{00}} = -\frac{L(LK')'}{K}  -
\left(\frac{KLN^{\phi'}}{2N}\right)^2  =  \Lambda +
\frac{\pi}{2}({\tilde{B}}^2 - {\tilde{E}^2})\eqno(2.20)$$

$$ \tilde{G_{11}} =  \frac{L(LN')'}{N}  -
3\left(\frac{KLN^{\phi'}}{2N}\right)^2  =  -\Lambda +
\frac{\pi}{2}
({\tilde{B}}^2 + {\tilde{E}}^2) \eqno(2.21)$$

$$\tilde{G _{01}} =  -\frac{L}{K^2} \left(\frac{K^3
LN^{\phi'}}{2N}\right)'  = -\pi \tilde{B}\tilde{E}  \eqno(2.22)$$

$$ \tilde{G_{22}} =  L^{2} \left( \frac{N'K'} {NK} \right) +
\left(\frac{KLN^{\phi'}}{2N}\right)^2  =  -\Lambda +
\frac{\pi}{2}(\tilde{B}^2 - \tilde{E}^2) \eqno(2.23)$$

\section{Exact Solutions}
We will look for exact solutions in which the electromagnetic
field is self- (antiself-)dual. Then we will have from (2.10)

$$ \tilde {E}  = \epsilon \tilde {B} = u(r)\eqno(3.1) $$
where
$\epsilon = 1(-1)$ corresponds to self (anti-self) duality
condition.
Then from (2.13), the non-zero components  of the
electromagnetic field in the  coordinate basis can expressed in
the form

$$E = F_{tr} = \frac{ u(N -\epsilon N^\phi K)} {L}$$
$$B = F_{r\phi} = \epsilon \frac {uK}{L}\eqno(3.2)$$

To solve the field equations, let us first define a function
$\rho(r)$ such that

$$ \frac{1}{L}= \frac{d\rho}{dr}\eqno(3.3)$$
Then, using (3.2), the
field
equations (2.5) can be solved to get,

$$u(\rho) =  C_{1} \frac {\exp[\epsilon m_{p}\rho]} {K}$$

$$ C_1\left( - N^{\phi} + \epsilon \frac{N}{K}\right) = \epsilon
C_{2}\eqno(3.4)$$

Using the gravitational field  equations (2.20) - (2.23), and the
conditions given by (3.4), we find that a general stationary,
circularly symmetric solution to the Einstein-Maxwell-ChernSimons
theory subject to (anti)self-duality condition (3.1) is
given by

$$\frac{1}{L} = \frac{d\rho}{dr}\eqno(3.5)$$.
$$K^{2} =  A_1 + A_2 \exp[2|\Lambda|^{1/2}\rho] - r_m^2
\exp[2\epsilon m_{p}\rho].\eqno(3.6)$$

$$N =  C_{0}\frac{\exp[2|\Lambda|^{1/2}\rho]}{K}\eqno(3.7)$$

$$N^{\phi} = \epsilon(\frac{N}{K} -
\frac{C_{2}}{C_{1}})\eqno(3.8)$$

Here, $$r_m^2 = \frac{\pi C_{1}^2 }{2m_{p}(m_{p} -\epsilon
|\Lambda|^{1/2})}\eqno(3.9)$$
and $C_{0}$,$C_{1}$,$C_{2}$, $A_{1}$ and $A_{2}$ are
integrating
constants. At this point, the solution depends on seven
parameters two of which are the topological mass and the
cosmological constants. We will see below how various physical
requirements can be used to reduce this number to three. We note
that these
solutions  represent the most general
solutions of our field equations except for the special cases 
$m_{p} = 0$ and
$m_{p} =  \epsilon |\Lambda|^{1/2}$.

These solutions is that they
determine the functions $L(r)$ and $K(r)$ only up to a choice of
parametrization (of coordinates). So, we will study these
solutions subject to the
additional requirement 
$K=r$. Moreover, to simplify the notation
and without loss of generality, we set $m_p = -\epsilon
|\Lambda|^{1/2}$. This
choice also leads to a potential for the electromagnetic field
with
interesting properties
which will be highlighted in a later section.
Under these conditions, we have, from  (3.6) 

$$r^{2} =  A_1 + A_2 \exp[2|\Lambda|^{1/2}\rho] - r_m^2
\exp[-2|\Lambda|^{1/2}\rho].\eqno(3.10)$$
We can solve this expression for $\rho$ to get

$$\rho = \frac{1}{2|\Lambda|^{1/2}} ln[X]\eqno(3.11)$$
where 

$$X = \frac{1}{2A_2} \left[r^2 - A_1 + \sqrt{(r^2 - A_1)^2  +
4A_2r_m^2}\right]\eqno(3.12)$$
So, the functions representing the  stationary charged solution 
are given by the following expressions :

$$L =  \frac{|\Lambda|^{1/2}}{r} \sqrt{\left[(r^2 - A_1)^{2}  +
24A_2r_m^2 \right]}\eqno(3.13)$$

$$N = \frac{C_o|\Lambda|^{1/2}}{2A_2r} \left[r^2 - A_1 +
\sqrt{(r^2
- A_1)^2  + 4A_2r_m^2 }\right]\eqno(3.14)$$

$$N^{\phi} = \epsilon \left(\frac{N}{K} - |\Lambda|^{1/2}\right)
\eqno(3.15)$$
where
$$r_m^2 = \frac{\pi C_{1}^2 }{4|\Lambda|}\eqno(3.16)$$
To see the significance of the constant $C_{1}$, 
we note that for large $r$ and in the limit of vanishing
topological mass, 
the electric and the magnetic fields, which are given by
$\tilde{E}
= \epsilon \tilde{B} = C_{1}\exp[\epsilon m_p\rho] /r$ approach
the flat space space expression 
$C_{1} /r$. Therefore, the constant $C_{1}$ can be
identified with the electric (magnetic) charge $Q_{e}$. Moreover,
requiring that the shift function $N^{\phi}$ remain finite as $r
\rightarrow \infty$, we must choose the constant
$C_{2}$ as follows : 
$$C_{2} = C_{1}|\Lambda|^{1/2}\eqno(3.17)$$ 

In the limit $Q_e \rightarrow 0 $, the above solution reduces to

$$L =  |\Lambda|^{1/2}\frac{(r^2 - A_1)}{r}\eqno(3.18)$$

$$N = \frac{C_o |\Lambda|^{1/2}}{2A_2} \frac{(r^2 -
A_1)}{r}\eqno(3.19)$$

and $$r^2 = A_1 + A_2 (r^2 - A_1).\eqno(3.20)$$
To satisfy equation (3.20), we must set $A_{2} = 1$. We are thus
left with two undetermined parameters $C_{0}$ and $A_{1}$. 
If we wish to cast these solutions 
in the BTZ form, we must choose $C_{0} = 1$.

Then the only parameter left to be determined is $A_{1}$. We will
see below that it can be expressed in terms of asymptotic values
of mass or angular momentum. To demonstrate this, we shall use
the quasilocal formalism developed by Brown, York,
Creighton and  Mann \cite{bro}\cite{man}.

\section{Quasilocal Mass and Angular Momentum}
In this section we will briefly review the formalism developed by
Brown {\it et al.} \cite{bro}\cite{man}, for computing the
expressions for quasilocal
energy and other conserved charges of a gravitating system. 
The conserved quasilocal charges such as the energy, $E$, are
defined as the total charges 
within a region of space with boundary, $B$, whenever that
boundary
admits a corresponding Killing
vector.
The conserved charge associated with a timelike Killing vector
defines
the quasilocal Mass, $M$, and the conserved charge associated
with the rotational
Killing vector defines the quasilocal Angular momentum, $J$.

Let the spacetime manifold under consideration be $\cal M$. We
take the
topology of $\cal M$ to be the product of a spacelike
hypersurface and a real line
interval, i.e., $\Sigma \times I$. The boundary of $\Sigma$ is
then given by
$B$.
The boundary of $\cal M$ , denoted by $\partial \cal M$, consists
of initial
and final 
spacelike hypersurfaces $t'$ and $t''$, respectively, and a
timelike
hypersurface
$\cal T$ = $B \times I$ joining the space-like hypersurfaces.
The induced metric on the hypersurfaces $t'$  and $t''$ is
denoted
by $h_{ij}$, and the induced metric  on $\cal T$ is defined by
$\gamma_{ij}$.
The boundary element $\cal T$ is foliated into one dimensional
hypersurfaces $B$ with the induced parameter $\sigma$. The
two-metric $\gamma_{ij}$ on $\cal T $ can be decomposed, using
the notation of equation (2.6), as 
follows: 

$$ \gamma_{ij}dx^idx^j = -N^2dt^2 + r^2(d\phi + N^{\phi}dt)(d\phi
+ N^{\phi} dt)\eqno(4.1)$$
Then on B the proper energy surface density $\varepsilon$ and
proper
angular momentum surface density $j_a$, on $B$ could be defined
as
follows \cite{bro}\cite{man}:

$$\varepsilon = \frac{k}{k_o} - \varepsilon_o \eqno(4.2)$$

$$j_{i} = \frac{-2}{\sqrt{h}} \sigma_{ij}P^{jk}n_{k} -
(j_{o})_i\eqno(4.3)$$
In these expressions,
$k$ is the  extrinsic curvature of $B$ considered as the boundary
$B = \partial \Sigma $. For our metric, it has the value $-L/r$. 
The quantities $P^{ij}$ denote the  gravitational
momentum conjugate to $h_{ij}$ which are, in turn, the components
of metric on the
spacelike hypersurfaces $t'$ and $t''$. In our case, the only
non-zero
component of $P^{ij}$ 
is $P^{r \phi}= -rLN^{\phi'}/ 4 \pi N$. The quantity $\vec{n}$ is
the unit normal to  $\cal T$ with  $n_{\mu} =   L
\delta^{1}_{\mu}$ for the given metric.  Also $\sigma$
= $r^2$, and  $h = Det(h_{ij})$ = $L^{-1}r$.

The values $\varepsilon_o$ and  $j_o$ corresponds to the ``zero
point configuration'' which arises due to the freedom of choice
in
the definition of $E$, $M$, and $J$. In the (2+1) dimensions, the
zeropoint configuration can be chosen to be, e.g., a stationary
slice of
the zero mass-black hole solution.

With these preliminaries, the total quasilocal energy for the
system is defined by,

$$ E = \int_B dx \sqrt{\sigma} \varepsilon \eqno(4.4)$$
For other charges, when there is a Killing vector field ${\bf
\xi}$ in $\cal T$, the corresponding
conserved charge is given by

$$Q_{\xi} = \int_B dx \sqrt{\sigma} ( \varepsilon u^i +
j^i)\xi_{i}\eqno(4.5)$$
The vector $\vec{u}$ is the unit normal to $\Sigma$ , with $u_\mu
= - N \delta^{0}_{\mu}$.
Thus, for a system with rotational (circular) symmetry and the
Killing vector field $\xi$ in $\cal T$, the conserved charge is
the angular momentum

$$ J = \int_B dx \sqrt{\sigma}j_i\xi^i\eqno(4.6)$$
Similarly, for the stationary spacetimes considered in this
paper, the
conserved anti-de Sitter mass, $M$, corresponding to the timelike
Killing vector
has the form

$$ M = \int_B dx \sqrt{\sigma} ( N\varepsilon - N^{\phi}j_i
)\eqno(4.7)$$

Using the above formalism , it is easy to show that for the
metric (2.6), with $K^{2} = r^{2}$,
the quasilocal angular momentum at radial distance $r$ is given
by

$$ J(r) = \frac{LN^{\phi'}r^3}{N}.\eqno(4.8)$$
Using equations (3.14)-(3.17), this takes the form

$$J(r)= 2|\Lambda|^{1/2}\epsilon \left[r^2 -
\sqrt{(r^2 - A_1)^2  + 4r_m^2 }\right].\eqno(4.9)$$
In the limit $r \rightarrow \infty$, we obtain the asymptotic
observable $J$ given by

$$J =J(\infty) = 2|\Lambda|^{1/2}\epsilon A_1 \eqno(
4.10)$$
Similarly, using the same formalism \cite{bro}, one can compute
the
expressions for quasilocal mass and quasilocal energy at the
radius $r$. The result is

$$M(r) = 2NE(r) - J(r)N^{\phi} 
\eqno(4.11)$$

$$E(r) =  \left[ L_{0} - L\right] \eqno(4.12)$$

The asymptotic observables $J$ and $M$, correspond to the Casimir
invariants of the anti-de Sitter group. Since there are only two
such invariants, it is clear that the quantity $E$ cannot be an
anti-de Sitter invariant. The significance of this quantity
becomes clear if we perform a group contraction on anti-de Sitter
to obtain the Poincare' group. In that case, looking at the
square
of the quantity $M(r)$ given by equation (4.11), we can see that
$E^{2}$ will be the only part of $M^{2}$ left over after group
contraction. So, it has the significance of mass (or energy) in
the sense of Poincare' group. We also note in passing that the
second Casimir invariant of the anti-de Sitter group does not
change under group contraction. Thus the notion of ``spin'' as
the square root of the eigenvalue of the second Casimir operator
has the same meaning in the two groups.
  
In equation (4.12), the quantity  $L_{0}$ is the reference value
of the function $L(r)$ which determines
the
zero of the energy. Such a reference spacetime can be chosen  by
setting certain integration constants in the solution to zero. If
we choose  $Q_{e}$ =0, M = 0,  the background metric approaches 

$$ds^2 = - |\Lambda|r^2 dt^2 +  \frac{dr^2}{|\Lambda |r^2} +
r^2d\phi^2\eqno(4.13)$$
This is identical to the vacuum anti-de Sitter spacetime or the
``zero mass black
hole configuration'' given in \cite{banados}. For such a
reference spacetime,

$$L_{0} = |\Lambda|^{1/2}r.\eqno(4.14)$$
Then, using (4.10) and (4.12), the expression for quasilocal
anti-de Sitter mass
takes the form
 $$ M(r) =  {2|\Lambda|}\left[r^2 - \left((r^2 - A_1)^2  + 4r_m^2
\right)^{\frac{1}{2}}\right].\eqno(4.15)$$
Again in the limit $r \rightarrow \infty$, we get the asymptotic
observable $M$:
$$M = M(\infty) = {2|\Lambda|A_1}\eqno(4.16)$$
This determines the last of our integration constants to be

$$A_1 = \frac{M}{2|\Lambda|}\eqno(4.17)$$
Then,

$$ J = \frac{\epsilon M}
{{|\Lambda|}^{1/2}}\eqno(4.18)$$
 
\section{\ Rotating Charged Solutions}
Compiling the results obtained in the previous sections, we now
present a family of rotating charged solutions for
Einstein-Maxwell theory in which the photon has a ChernSimons
mass term. They are characterized by mass
$M$, the  angular momentum $J$, and
charge $Q_e$. The spacetime metric of
these solutions has the form,

$$ds^2 = -N^2 dt^2 + L^{-2}dr^2 +  r^2 ( d\phi +
N^{\phi}dt)^2.\eqno(5.1)$$
where

$$L^2 =  \frac{|\Lambda|}{r^2}\left[\left(r^2 -
\frac{M}{2|\Lambda|}\right)^2  + \frac{\pi
Q_e^2}{|\Lambda|}\right].\eqno(5.2)$$

$$N = \frac{|\Lambda|^{1/2}}{2r} \left[r^2 - \frac{M}{2|\Lambda|}
+ \sqrt{(r^2 - \frac{M}{2|\Lambda|})^2 + \frac{\pi
Q_e^2}{|\Lambda|}}\right].\eqno(5.3)$$

$$N^\phi = \frac{J}{2r^2} \left[ \frac{|\Lambda|}{M}\sqrt{(r^2 -
\frac{M}{2|\Lambda|})^2  + \frac{\pi Q_e^2}{|\Lambda|^2}} -
\frac{r^2|\Lambda|}{M} - \frac{1}{2}\right].\eqno(5.4)$$
These solutions are regular and horizonless. They have
finite mass $M$ and finite angular momentum $\epsilon M
|\Lambda|^{-1/2}$. Since the electromagnetic fields are
(anti)self-dual, the
trace of the energy momentum tensor $T^{\mu}_{\mu} =
F^{\mu\nu}F_{\mu\nu} = 0$, leading to the same constant negative
scalar curvature  $R = 6\Lambda$ as for the uncharged BTZ black
hole solution. Thus, locally,
the space-time geometry would have an anti-de Sitter structure.

The electromagnetic potential for these solutions is given by,

$$A_{\mu} dx^{\mu} = \frac{Q_e}{\sqrt{2}} \left[r^2 -
\frac{M}{2|\Lambda|} + \sqrt{(r^2 - \frac{M}{2|\Lambda|})^2 + 
\frac{\pi Q_e^2}{|\Lambda|}}\right]^{-1/2}\left(
|\Lambda|^{-1/2}d\phi -dt\right).\eqno(5.5)$$
Thus for large $r$,
 the Electromagnetic potential behaves as

$$A_{\mu}(r) \rightarrow r^{-1}\eqno(5.6)$$
This implies that the field strengths behave asymptotically as

$$\tilde{E} = \epsilon \tilde{B} \rightarrow  r^{-2}    
\eqno(5.7)$$
It is interesting to note that the potential in (5.6) 
has the behavior of the Coulomb field in flat space in $3+1$
dimensions.
In fact, had we not related the topological mass to the
cosmological constant, i.e., for arbitrary $m_p$, the potential
would behave as 
$r^{\frac{\epsilon m_p}{|\Lambda|^{1/2}}}$. We may recall that in
the absence of mass term, the electromagnetic potential in $2+1$
dimensions behaves logarithmically.  In our solutions, by
introducing a 
topological mass term which behaves as an infrared regulator, 
we have been able to modify the long range behavior of the
electromagnetic potential. In flat space, the addition of the
topological 
mass term will result in an exponential modification of the field
strengths :

$$\tilde E = \epsilon \tilde B = Q_e \frac{e^{\epsilon
m_{p}r}}{r}\eqno(5.8)$$ 
In curved
space, the effect is more subtle and, of course, coordinate
dependent.

It is also interesting to note that when  $Q_e \rightarrow 0$,
the our charged 
solutions are asymptotic to the extreme uncharged BTZ blakchole
with $|J| = M
|\Lambda|^{-1/2}.$ 
To see this, we note that for 
$\frac{M}{2|\Lambda|} >> Q_e^2$, and large r, 

$$N = L \rightarrow  |\Lambda|^{1/2}\frac{(r^2 -
\frac{M}{2|\Lambda|})}{r}\eqno(5.7)$$
$$ N^{\phi} \rightarrow  \frac{-J}{2r^2}\eqno(5.8)$$
These expressions are identical to the corresponding expressions
for
the extreme uncharged
BTZ black
hole solution with mass $M$ and angular momentum $J$.

\section{Comparison with previous charged solutions.}
The static charged black hole solution  discussed in
\cite{banados}, was specified by three parameters : the mass
$M$, the charge $Q_e,$, and the `` radial parameter'' $r_o$.
Depending on the values of these parameters, a charged BTZ
solution can have
one, two, or no horizons. The charged solution with one horizon
was identified as an extreme charged black hole. However,
the
quasilocal mass for this solution is given by 
$$M_{ql} = M + Q_{e}^2 \ln[\frac{r}{r_0}]$$
In contrast to what happens in our solutions, this quantity
diverges for
large $r$ \cite{eric}.
Consider next the rotating charged black hole solution for self-
(antiself-) dual
Einstein-Maxwell presented by Kamata {\it et al.} \cite{kam}.
This
solution has a  horizon at $r = r_{0}$. Its angular momentum and
the quasilocal mass for large $r$ were computed by
Chan\cite{chan}:

$$j(r) = \frac{2Q_{e}^2}{|\Lambda|}\ln |\frac{r^2 -
r_0^2}{r_0^2}|\eqno(6.1)$$

$$M(\infty) = 2 |\Lambda|\left( \frac{Q^2}{\Lambda} +
r_{0}^2\ln |\frac{r^2}{r_0^2}|\right)\eqno(6.2)$$
They both diverge for large $r$.
In both of these solutions, the divergences are  due to the
logarithmic behavior of the electromagnetic potential in 2+1
dimensions as we mentioned before. In the our solutions, the
``screening'' effect of the topological mass term is responsible
for the finite values
of mass and angular momentum.

After the completion of this work, we learned that 
using a dimensional reduction method, general 
solutions for Einstein-Maxwell-Chern Simons theory equivalent to
ours have also been obtained by Cl\'ement \cite{clem}. Since the
method of obtaining these solutions in Cl\'ement's work are
completely independent of our methods, the two works together
provide a more convincing proof of the generality of these
solutions. This is, in particular, evident in our direct
computation of the masses and angular momenta.

\bigskip
This work was supported in part by the Department of Energy under
contract number DOE-FG02-84ER40153.

\end{document}